\pdfoutput=1
\documentclass[11pt]{article}
\usepackage[colorlinks=true, pdfstartview=FitV, linkcolor=blue, citecolor=blue,urlcolor=blue]{hyperref}

\usepackage{amsbsy}
\usepackage{amssymb}

\usepackage{verbatim}

\makeatletter
\@addtoreset{equation}{section}
\makeatother

\newtheorem{theorem}{Theorem}[section]

\newtheorem{exercise}{Exercise}[section]

\newtheorem{lemma}{Lemma}[section]
\newtheorem{remark}{Remark}[section]

\newtheorem{proposition}{Proposition}[section]
\newtheorem{corollary}{Corollary}[section]
\newtheorem{definition}{Definition}[section]
\def\le{\left}
\def\ri{\right}
\def\ds{\displaystyle}

\def\res{\mathop{\mathrm {res}}\limits_}

\def\br{\begin{remark}}
\def\er{\end{remark}}
\def\bt{\begin{theorem}}
\def\et{\end{theorem}}
\def\bc{\begin{corollary}}
\def\ec{\end{corollary}}
\def\bx{\begin{examp}\small}
\def\ex{\end{examp}}
\def\bxr{\begin{exercise}\small}
\def\exr{\end{exercise}}
\def\bl{\begin{lemma}}
\def\el{\end{lemma}}
\def\bd{\begin{definition}}
\def\ed{\end{definition}}
\def\bp{\begin{proposition}}
\def\ep{\end{proposition}}
\def\be{\begin{equation}}
\def\ee{\end{equation}}
\def\&{\hspace{-15pt}&}
\def\bea{\begin{eqnarray}}
\def\eea{\end{eqnarray}}
\def\beas{\begin{eqnarray*}}
\def\eeas{\end{eqnarray*}}

\def \pa{\partial}
\def\C{{\mathbb C}}

\def\R{{\mathbb R}}
\def\N{{\mathbb N}}

\def\wh{\widehat}

\def\Z{{\mathbb Z}}

\def\a{\alpha}
\def\d{\,\mathrm d}

\def\1{{\bf 1}}
\def\wt{\widetilde}

\date{}
\begin{document}
\baselineskip 15pt plus 1pt minus 1pt

\vspace{0.2cm}
\begin{center}
\begin{Large}
\fontfamily{cmss}
\fontsize{17pt}{27pt}
\selectfont
\textbf{
Moment determinants as isomonodromic tau functions}
\end{Large}\\
\bigskip
\begin{large} {M.
Bertola}$^{\ddagger,\sharp}$\footnote{Work supported in part by the Natural
    Sciences and Engineering Research Council of Canada
(NSERC).}\footnote{bertola@crm.umontreal.ca}
\end{large}
\\
\bigskip
\begin{small}
$^{\ddagger}$ {\em Department of Mathematics and
Statistics, Concordia University\\ 1455 de Maisonneuve W., Montr\'eal, Qu\'ebec,
Canada H3G 1M8} \\
$^{\sharp}$ {\em Centre de recherches math\'ematiques, Universit\'e\ de
Montr\'eal } \\
\end{small}
\bigskip
{\bf Abstract}
\end{center}
We consider a wide class of determinants whose entries are moments of the so-called semiclassical functionals and we show that they are tau functions for an appropriate isomonodromic family which depends on the parameters of the symbols for the functionals. This shows that the vanishing of the tau-function for those systems is the obstruction to the solvability of a Riemann--Hilbert problem associated to certain classes of (multiple) orthogonal polynomials. 
The determinants include H\"ankel, T\"oplitz and shifted-T\"oplitz determinants as well as determinants of bimoment functionals and the determinants arising in the study of multiple orthogonality.
Some of these determinants appear also as partition functions of random matrix models, including an instance of a two-matrix model.

\vspace{0.7cm}

\vskip 15pt
{AMS-MSC2000: 05E35, 15A52}

\tableofcontents

\section{Introduction}

The connection between orthogonal polynomials and monodromy preserving deformations of differential equations has been observed repeatedly \cite{Magnus:Painleve,FIK0,FIK1,FIK2,Its:RandomWords2,Bertola:Semiiso,Bertola:LOPs}. It can be used to produce special solution to Painlev\'e\ equations by choosing appropriately the measure of orthogonality and its dependence on deformation parameters \cite{Magnus:Painleve}. 
In the eighties the notion of isomonodromic tau function was developed \cite{JMU1,JMU2,JMU3} for systems of monodromy preserving deformations of rational connections (under some genericity assumptions); the philosophical attitude was  that --in the context of Birkhoff--Riemann--Hilbert problems-- the tau-function should play the same r\^ole as the classical Theta function plays for the Jacobi inversion problem on algebraic curves. 
Indeed the vanishing of a Theta-function on the Abel map of a divisor determines its ``specialty'', namely the (im)possibility of finding suitable meromorphic functions with prescribed divisor. Similarly the vanishing of the  isomonodromic tau function indicates that a Riemann--Hilbert problem is not solvable. This was proved first for the Schlesinger isomonodromic equations by Malgrange \cite{Malgrange:IsoDef1, Malgrange:IsoDef2, Malgrange:Deformations} and more recently for connections with irregular singularities by Palmer \cite{Palmer:Zeros} (although under some genericity assumptions). 

In this perspective a connection between H\"ankel determinants and isomonodromic tau functions is not surprising since the vanishing of the H\"ankel determinant of the moments of a measure determines whether the orthogonal polynomials exist or not and  -on the other hand- their existence is equivalent to the solvability of a RHP \cite{FIK1}. 

The precise functional relationship between these (and similar) determinants and isomonodromic tau functions was established in \cite{Bertola:PartitionJPA, Bertola:Semiiso} for the H\"ankel case and in \cite{Its:RandomWords2,Bertola:LOPs} for shifted Toeplitz determinants of so-called {\em semiclassical moment functionals} \cite{Magnus:Painleve, Marcellan:Semiclassical1,Marcellan:Semiclassical2}.

In a seemingly distant area, moment determinants appear as partition functions for random matrix models \cite{Mehtamethod,MehtaBook}, which are known to be tau functions (in a different sense) for the KP, Toda or 2Toda hierarchies, depending on the case.
For example the partition function of the  Hermitean matrix model defined as 
\be
Z_n^{1MM} :=  \int_{\mathcal H_n} \d M {\rm e}^{-{\rm Tr} V(M)}
\ee
where $ M$ is a Hermitean matrix and  $\d M$ is the Lebesgue measure on the vector space $\mathcal H_n $ of Hermitean matrices of size $n\times n$; here $V(x)$ represents any scalar function, but for the sake of the example we can take it to be a polynomial.

By means of Dyson-Mehta theorem \cite{MehtaBook} it is known that $Z_n^{1MM}$ is (up to an inessential normalization constant) the same as a H\"ankel determinant for the moments of the measure (or weight) ${\rm e}^{-V(x)} \d x$
\be
Z_n^{1MM} \propto \Delta_n:= \det[\mu_{i+j}]_{0\leq i,j\leq n-1}\ ,\ \ \ \mu_j:= \int x^j {\rm e}^{-V(x)} \d x\label{1MM}
\ee
In addition the same quantity can be expressed as a multiple integral over the spectra 
\be
Z_n^{1MM} \propto \int\d x_1 \dots \int \d x_n \Delta^2 (X) {\rm e}^{-\sum_{j=1}^n V(x_j)}\ ,\ \ \Delta(X):= \prod_{j<k} (x_j-x_k) 
\ee
This correspondence between matrix models and H\"ankel determinants is much deeper, inasmuch as the correlation functions for the random eigenvalues can all be described in terms of the orthogonal polynomials for the above measure.
 
The relationship between $Z_n^{1MM}$ and isomonodromic tau functions was established in \cite{Bertola:Semiiso, Bertola:PartitionJPA} but in this paper we will provide a new proof which applies to more general cases as well.
An example is the following two--matrix model, recently introduced in \cite{Bertola:CauchyMM}; it consists of pairs $M_1,M_2$ of positive-definite Hermitean matrices of size $n\times n$  with a partition function defined as follows
\be
Z_{n}^{2MM}:= \int \int \d M_1 \d M_2 \frac{ {\rm e}^{-{\rm Tr} V_1(M_1) - {\rm Tr} V_2(M_2)}}{\det(M_1+M_2)^n}
\ee
It can be shown that this partition function is also a moment determinant \cite{Bertola:CauchyMM}
\be
Z_{n}^{2MM}:= \det [{\mathcal I}]_{jk}\ ,\ \ \ \mathcal I_{jk}:= \int_{\R_+} \int _{\R_+} \frac {{ \rm e}^{-V_1(x)-V_2(y)}\d x \d y}{x+y} x^j y^k
\ee
 Similarly to the Hermitean-matrix model case, also here the spectral statistics can be solved in terms of  certain biorthogonal polynomials for the pairing induced by the integrand above \cite{Bertola:CBOPs, Bertola:CauchyMM}

A third class of moment determinants that we will consider here and were considered in \cite{Bertola:LOPs}  are the shifted Toeplitz determinants 
\be
\Delta_n^\ell:= \det[\mu_{\ell + j-k} ]_{0\leq j,k\leq n-1}
\ee
for moments defined as in (\ref{1MM}). These enter as tau-functions for various integrable lattices of Toda/Ablowitz-Ladik type \cite{Gekhtman:InverseMoment,Gekhtman:Poisson,Gekhtman:Elementary}. In the ordinary Toeplitz case ($\ell =0$) they too can be interpreted as partition functions for unitary matrix models.

A last class of multi-weight moment determinant we will consider here arise in the theory of multiple Pad\'e\ approximation for Markov functions \cite{Aptekarev:Multiple}. Let ${\rm e}^{V_\ell(x)}\d x\ ,\ \ j=1,\dots, K $ be a collection of (semiclassical) weights with polynomial $V_\ell$'s (for the sake of this example).
For  a point $\vec n = [n_1,\dots, n_K]$ in the lattice $\N^K$ we define the following determinant of size $n= |\vec n| := \sum_{\ell=1}^K n_\ell$

\be
\Delta_{\vec n}:=  \det \le[
\begin{array}{ccccccc}
\mu^{(1)}_0 &&&&&& \mu_{n-1}^{(1)}\\
\vdots &&&&&& \vdots\\
\mu^{(1)}_{n_1-1} &&&&&&\mu_{n+n_1-2}^{(1)}\\
\hline
\vdots &&&&&\vdots\\
\hline
\mu^{(K)}_0 &&&&&& \mu_{n-1}^{(K)}\\
\vdots &&&&&& \vdots\\
\mu^{(K)}_{n_K-1} &&&&&&\mu_{n+n_K-2}^{(K)}\\
\end{array}
\ri]\ , \ \ \mu_j^{(\ell)} := \int x^j {\rm e}^{-V_\ell(x) } \d x
\ee
\br
We ignore whether there is a matrix model related to this determinant.
\er

The point of the mater is that for all these determinants,  the dependence on the coefficients of the potentials $V$ is {\em isomonodromic} (in the sense of Jimbo-Miwa-Ueno) and that they can all be identified with the isomonodromic tau function of \cite{JMU1}. The paper is devoted to this relationship.
\subsection{Organization of the paper and outline of the proof}
We first review very quickly the setup of \cite{JMU1,JMU2,JMU3} in Sec. \ref{RevIso}, in order also to establish the notation and  some results that  are  used later. We will need to compute explicitly the isomonodromic tau function for a very simple (almost trivial) class of isomonodromic families of upper-triangular matrices (Sec. \ref{sec_trivial})  since this result will be used repeatedly.
In Sec. \ref{SecMarkov} we recall in more detail the notion of semiclassical moment functional and associated Markov (Weyl) function.

The last four sections are devoted to establishing the precise relationship between determinants of type H\"ankel (Sec. \ref{Hankel}), shifted-Toeplitz (Sec. \ref{Toeplitz}), Cauchy/bimoment (Sec. \ref{Cauchy}) and multiple-weight (Sec. \ref{Multimeas}). 

We briefly comment on the simple idea of proof since it is the same in all cases with changes only in the details of the  implementation.

The first step is to consider the relevant (multi)orthogonal polynomials associated to the determinant under inspection and characterize them by means of a Riemann--Hilbert problem. Most part of this step is actually well known (for example \cite{FIK0,FIK1} for the case of orthogonal polynomials) and can be found --for the different cases-- in the literature.

The second step is to establish an ODE whose (generalized) monodromy is constant by construction under deformations of the weights; this is accomplished by transforming the RHP into an equivalent one with constant jumps. This step requires the weights to be semiclassical in order to obtain a rational ODE and thus to fall within the framework of isomonodromic deformations. At this point the isomonodromic tau function recalled in Sec. \ref{RevIso} is constructed. 
We then recognize that the changes in the degrees (or multi-degrees) of the polynomials are suitable Schlesinger transformations and thus we can prove that  the ratio of the corresponding tau function is the same as the ratio of the corresponding moment determinants.

Therefore to complete the description of the relationship it is sufficient to identify $\tau$ and the determinant for a ``trivial'' case (the ``initial condition''). This is always the case of polynomials  of degree $0$ and determinants of size $0$ (which equal $1$ by convention). The isomonodromic family corresponding to degree-zero (multi) orthogonal polynomials is --in all cases-- an instance of the ``trivial'' class of Sec. \ref{sec_trivial} for which the isomonodromic tau function can be computed easily and explicitly. 

\section{Isomonodromic deformations: a (short) review}
\label{RevIso}
The content of this section is lifted (with adapted notation) from \cite{JMU1, JMU2, JMU3} and we refer the reader  ibidem for more details.

Consider an rank-$r$ ODE in the complex domain 
\be
\Psi'(x) = D(x) \Psi(x)\ ,\ \ D(x)\in Mat_{r\times r}\label{ODEpsi}
\ee
 with an $r\times r$ matrix $D(x)$ whose coefficients are rational functions. We will denote by $\mathcal A$ the divisor of singular points (including $\infty$ if the case) and by $a\in \mathcal A$ a point in it; for $a\in \mathcal A$ we denote by $\zeta= \zeta_{.
a}$ the {\bf local parameter at $a$}, namely $\zeta_a = x-a$ if $a\neq \infty$ or $\zeta_\infty = \frac 1 x$ for $a=\infty$. We will denote by $d_a+1$ the order of the pole of $D(x)\d x$ at $x=a$. 

If the differential  $D(x) \d x$ has a pole at $a$ of order $\geq 2$ then the point is called an {\bf irregular singularity}\footnote{When counting poles for $D(x)\d x$ we must recall that at $x=\infty$ the differential $\d x$ has a second-order pole.}, otherwise it is called a {\bf Fuchsian singularity}.

Suppose that near a singularity $x=a$ we have the following (possibly sectorial) asymptotic expansion 
\be
\Psi(x) = G (\overbrace{\1 + Y_1\zeta + Y_2\zeta^2 + \dots}^{Y(\zeta)} ) {\rm e}^{T(\zeta)}\label{exppsi}
\ee
where $T(\zeta)$ is a {\em diagonal} matrix of the form 
\be
T(\zeta) = \sum_{j=1}^{d_a} T_j \zeta^{-j} + T_0 \ln \zeta 
\ee
It is understood that $T=T_{_{(a)}}, Y = Y_{_{(a)}}, G = G_{(a)}$ are matrices that depend on the particular singularity under exam; when necessary we will use a subscript to remind of this fact.
We recall the following fact:
\bp[\cite{Wasow}]
\label{suffic}
A {\em sufficient} condition for the expansion (\ref{exppsi}) to be valid in suitable sectors around an irregular singularity $x=a$ is that the leading coefficient of the singularity of $D(x)\d x$ is {\em ad-regular} namely it has simple spectrum (i.e. distinct eigenvalues). 
\ep
We stress --which is crucial for our setting-- that the above condition  is {\em only  sufficient} and not necessary; in fact some of our cases below will violate the condition but still retain the validity of the expansion (\ref{exppsi}).

The expansion  (\ref{exppsi}) is in general only asymptotic, namely the series $Y(\zeta)$ is only formal;  
a theorem \cite{Wasow} shows however  that for each (sufficiently small)  sector there is a {\em bona fide} solution $\Psi$ whose asymptotic expansion matches (\ref{exppsi}). Two solutions $\Psi,\wt \Psi$ having the {\em same} asymptotics in {\em different} sectors are related by a right constant multiplier $\Psi(x) = \wt \Psi(x) S$, called a {\bf Stokes matrix}.  Given two singularities $a,\wt a\in \mathcal A$ and suitable sectors near them, the right multiplier for the solutions $\Psi, \wt \Psi$ with the given asymptotics is called {\bf connection matrix}.

Additionally, analytic continuation of any solution $\Psi$ to the punctured plane $\C\setminus \mathcal A$ induces a (anti) representation of the fundamental group called the {\bf monodromy representation} for the connection $(\pa_x -D(x)) \d x$.

We will assume that $\infty$ is a singularity and choose the normalization of $G_\infty=\1$ (this can always be accomplished by a left-multiplication of $\Psi$ by  constant invertible matrix). 

\bd
The generalized monodromy data consist of the totality of the Stokes and connection matrices, the monodromy representation and the $T_{0,a}$'s appearing in (\ref{exppsi}). 
\ed
\br
This definition is redundant since the data as defined above are not entirely independent or ``minimal''. This is of no consequence since we are going to study only transformations that preserve those data.
\er

\bd
The {\bf isomonodromic times} are the coefficients $T_{j,a}\ ,\ j\geq 1\ \forall a\in \mathcal A$  as well as the location of the singularities.
\ed
In \cite{JMU1} it was shown how to deform the isomonodromic times while preserving the generalized monodromy data of the connection. 

\bd
The {\bf isomonodromic tau function} $\tau$ is a function of the isomonodromic times (depending parametrically on the generalized monodromy) defined via the following first-order PDE
\be
\delta \ln \tau := \sum_{a\in \mathcal A} \res{x=a} {\rm Tr} \le( Y_{_{(a)}} ^{-1}(x) Y'_{_{(a)}}(x) \delta T_{_{(a)}}(x)\ri)\d x\label{isotau}
\ee
where the symbol $\delta$ stands for the total differential in all isomonodromic times.
\ed
Of course closedness of the differential (\ref{isotau}) needs to be proved, and that was accomplished in \cite{JMU1}.

We point out that in \cite{JMU1} they were working on the assumption that all singularities of $D(x)$ met the sufficient condition of Prop. \ref{suffic}; however it can be seen --going through the proofs-- that the only information that they used was the validity of the expansion (\ref{exppsi}) with a diagonal $T_{(a)}(\zeta)$ near each singularity $a\in \mathcal A$

\subsection{``Trivial'' isomonodromic systems}
\label{sec_trivial}
We will need the explicit expression for $\tau$ in a very simple situation. Of course the simplest situation would correspond to a diagonal connection, for the problem would be immediately solvable. 
Suppose that $D(x)$ is upper-triangular and also that  at all singularities $a\in \mathcal A$ we have 
\be
\Psi(x) \sim Y_{_{(a)}}(x) {\rm e}^{T_{_{(a)}}(x)}
\ee
where $T_{_{(a)}} (x)$ is diagonal and $Y_{_{(a)}} (x)$ upper triangular and (formally) analytic in the local parameter $\zeta = \zeta_{_{(a)}}$.
Denoting by $Y_d(x)$ the diagonal part of $Y$ the isomonodromic tau function is simply given by 
\be
\d \ln \tau  = \res{} {\rm Tr} (Y_d^{-1}Y_d' \d T)
\ee
For definiteness let 
\be
D(x) = -\le[\begin{array}{ccc}
V_1'(x) & \star & \star\\
&\ddots &\\
&& V_r'(x)
\end{array}\ri]
\ee
with $V_j'(x)$ some rational functions.
Then the solution has the following form 
\be
\Psi = \le[
\begin{array}{ccc}
{\rm e}^{-V_1} &\star &\star\\
& \ddots &\star\\
&& {\rm e}^{-V_r}
\end{array}
\ri]
\ee
Near a singularity $x=a$ we thus have that $T(x) = -{\rm diag} (V_1,\dots, V_r)_{sing}$ is the {\em singular} part of the potentials at $x=a$ whereas $Y_d(x) = \exp \le(-{\rm diag} \le[ V_1,\dots, V_r\ri]_{reg}\ri)$ where the subscript denotes the regular part. 

Thus we have 
\be
\delta \ln \tau = \sum_{j=1}^r \res{a\in \mathcal A} (V_j')_{a, reg} \delta (V_j)_{a, sing}
\ee
where the symbol $\res{\mathcal A}$ denotes the sum of the  residues at all singularities.

A simple exercise shows that 
\be
\ln\tau = \sum_{j=1}^r \res{a\in \mathcal A} (V_j)_{a, reg} (V'_j)_{a,sing}.
\label{lepalle}
\ee

It should be noted that there could be other isomonodromic parameters in the upper triangular terms of $D(x)$ (and hence of $Y(x)$), but clearly the tau function does not depend on them. 
 
\subsection{Schlesinger transformations}
\label{Schlesinger}
The family of monodromy-preserving transformations can be slightly enlarged by allowing {\em discrete} transformations that change the entries of $T_{0,(a)}$ by {\em integers}
\be
T_{0,(a)} \mapsto T_{0,(a)} + L_{_{(a)}}\ ,\qquad L_{_{(a)}} = {\rm diag}( \ell_{(a),1} \dots \ell_{(a),r})\ ,\qquad \ell_{(a),j} \in \Z.
\ee
subject to the constraint $\sum_{a\in \mathcal A} {\rm Tr} L_{_{(a)}} = 0$. These transformations can be achieved by multiplying on the left the solution $\Psi$ by a {\em rational} matrix $R(x)$. Such matrix can be algorithmically constructed from the entries of the matrices $\{G_{_{(a)}}, Y_{_{(a)}}\}_{a\in \mathcal A}$ 
as explained in \cite{JMU2}. They are called {\bf Schlesinger transformations} and they do not change (by definition) neither the monodromy representation nor  the Stokes and connection matrices.
Schlesinger transformations form a discrete Abelian group (a lattice) of  transformations generated by the so--called {\bf elementary Schlesinger} transformations. These are such that the matrices $\{L_{_{(a)}}\}_{a\in \mathcal A}$ have only two nonzero entries, a $+1$ and a $-1$. For example one may have $L_{_{(a)}} = {\rm diag} (1,0,0\dots)$ and $L_{(\infty)} = {\rm diag}(0,0,-1,0,\dots)$.

The details can be found in \cite{JMU2}, but to give a taste of the type of computation we consider the elementary Schlesinger transformations where $L_{(\infty)} = {\rm diag}(0,\dots,0,1, 0,\dots,0,-1,0,\dots) = \sigma_{ij}$, where the $1$ is on the $i$-th position and the $-1$ is on the $j$-th.

Denote $Y_{(\infty}) (x) = Y(x) = \1 + \sum_{j=1}^{\infty} Y_j x^{-j}$; the problem is then to find a second formal series   $\wt Y(z) = \1 + \sum_{j=1}^{\infty} \wt  Y_j x^j$ and a (polynomial) matrix $R(x)$ such that 
\be
\wt Y(x)x^{\sigma_{ij}} =  R(x) Y(x) \label{sigmaij}
\ee
The solution is straightforward  but tedious: $R(x)$ must be linear in $x$ and of the form $R(x) = R_0 + xR_1$.
Matching the coefficients of the powers of $z$ carefully one can find $R_0$ and $R_1$ 
\be
R (x)  = \le[
\begin{array}{c|c|c|c|c}
\matrix{1&&\cr
&\ddots&\cr
&&1} & \matrix{a_1 \cr\vdots\cr a_{i-1} } &  &  & \\
\hline
\matrix{b_{1}  \dots b_{{i-1}}} & \rule{0pt}{15pt}x+A &  \matrix{b_1 & \dots &b_{j-1}} & b_j &\matrix{b_{j+1}  \dots  b_{r}}\\
\hline
& \matrix{a_{i+1} \cr\vdots\cr a_{j-1} } &\matrix{1&&\cr
&\ddots&\cr
&&1} & &\\
\hline 
&\frac 1{b_j}  \rule{0pt}{15pt}& &0 &\\  
\hline 
& \matrix{a_{j+1} \cr\vdots\cr a_{r} } &&&\matrix{1&&\cr
&\ddots&\cr
&&1} 
\end{array}
\ri]
\ee
where 
\be
A:= \frac {\ds - Y_{2,_{ij}} +  \sum_{\ell\neq i} Y_{1,_{i,\ell}}Y_{1,_{\ell,j}}}{Y_{1,_{ij}}}\ ,
\ \ 
a_\ell = -\frac {Y_{1,_{\ell j}}} {Y_{1,_{ij}}}\ ,\ \ 
b_\ell = - Y_{1,{i \ell}}
\ee
It appears quite obviously that 
\bp
The problem above admits solution if and only if $Y_{1,_{ij}} \neq 0$.
\ep

The importance for us lies in the following theorem --which we paraphrase--
\bt[Thm. 4.1 in \cite{JMU2}]
\label{tauratio}
Given two connections $D(x)\d x$ and $\wt D(x) \d x$ related by an elementary Schlesinger transformation  with a $+1$ in the $i$-th diagonal element of $L_{_{(a)}}$ and a $-1$ in the $j$-th one of $L_{(a')}$  then 
the corresponding isomonodromic tau functions $\tau, \wt \tau$ are related by 
\be
\delta \ln \le(\frac {\wt \tau}{\tau}\ri)
 = \delta H
\ee
where 
\be
H = \le\{
\begin{array}{ll}
(Y_{(a),1})_{ij} & \hbox { if } a=a'\\
(G_{(a')})_{ij} & \hbox { if } a=\infty,\ a'\neq \infty\\
(G_{_{(a)}}^{-1} )_{ij} & \hbox { if } a\neq\infty,\ a'= \infty\\
\frac{(G_{_{(a)}}^{-1} G_{(a')})_{ij}}{a'-a} & \hbox { if } a,a' \neq \infty,\ a'\neq a
\end{array}
\ri.
\label{taudets}
\ee
\et

With these preparations we are ready to investigate the relationship between several types of moment determinants and isomonodromic tau functions.
\section{Markov functions for semiclassical moment functionals} 
\label{SecMarkov}
We will consider a rather wide class of determinants of matrices whose entries are moments of (collections of) weights.

The weights we are considering are all of the {\em semiclassical type} as defined in \cite{Magnus:Painleve, Marcellan:Semiclassical1, Marcellan:Semiclassical2, Bertola:Semiiso, Bertola:BilinearJAT}. This  means that they are of the form $\mu(x) = {\rm e}^{-V(x)}$ with $V'(x)$ an arbitrary rational function.

They are integrated over contours $\gamma_j$ which can be arbitrary contours in the complex plane as long as  all integrals $\int_{\gamma_j} x^k \mu(x) \d x $ are convergent integrals. The range for $k$ will be either $\N$ or $\Z$, depending on the situation; a detailed description of the contours can be found in \cite{Bertola:Semiiso, Bertola:LOPs} and we refer thereto for a more detailed discussion.

We will choose arbitrary complex constants $\varkappa_j$ for each contour $\gamma_j$ and use the notation 
\be
\int_\kappa x^k \mu(x) \d x := \sum \varkappa_j \int _{\gamma_j} x^k \mu(x) \d x  = \mu_k
\ee
We will also use the notation $\kappa: \C \to \C$ to indicate  the locally constant function that takes the (constant) value $\varkappa_j$ on the corresponding contour $\gamma_j$.  The Markov function (sometimes referred to as Weyl function) for these semiclassical weights is simply defined as the locally analytic function on $\C \setminus \cup \gamma_j$ given by 
\be
W(x):= \int_\kappa \frac{ \mu(\zeta) \d \zeta}{\zeta -x}\  ,\qquad
\kappa := \sum_j\varkappa_j  \chi_{\gamma_j}(x)
\ee
At times we will denote by $\kappa$ (by abuse of notation) the {\bf support} of $\kappa$.
The function $W(x)$ has logarithmic growth at the {\bf hard--edges}, namely endpoints of contours $\gamma_j$ where $\mu$ is $\mathcal O(1)$. In this case one verifies that $W(x) = \mathcal O(\ln |x-a|)$, where $a$ is the hard--edge point.

\section{H\"ankel determinants: orthogonal polynomials}
\label{Hankel}
\subsection{Riemann-Hilbert problem of orthogonal polynomial}
Let $\{p_n(x)|n=0,1,2,...\}$ be the  (monic) OPs that satisfy the following orthogonality
\be\label{orthogonality}
\int_{\varkappa}  p_n(x) p_m(x)  {\rm e}^{-V(x)}\d x = h_n\delta_{nm}.
\ee

They can be uniquely characterized by  the Riemann-Hilbert Problem (RHP) described hereafter.
Define for $z\in \C \setminus \R_+$ the matrix
\be
\Gamma(z):= \Gamma_n(z):= \le[
\begin{array}{cc}
p_n(z) & {\cal C}[p_n](z)\\
\frac {-2i\pi}{h_{n-1}}  p_{n-1}(z) & \frac {-2i\pi}{h_{n-1}}{\cal C}[p_{n-1}](z)
\end{array}
\ri]\ ,\qquad {\cal C}[p](z):= \frac 1 {2i\pi} \int_{\varkappa} \frac {p(x) {\rm e}^{- V(x)}\d x}{x-z}.
\label{OPRHP1}
\ee
The above matrix has the following jump-relations and asymptotic behavior that
uniquely characterize it \cite{FIK0, FIK1, FIK2, FIK3} (we drop the explicit dependence on
$n$ for brevity)
\bea
\Gamma_+(x) = \Gamma_{-}(x) \le[
\begin{array}{cc}
1 & \varkappa_j  {\rm e}^{-V(x)}\cr
0&1
\end{array}
\ri]\ ,\ \ x\in\Gamma_j\ ,\qquad
\Gamma(z) \sim\le(\1 + \sum_{\ell=1}^\infty \frac {Y_\ell} {z^\ell}\ri) \le[\begin{array}{cc}
z^n &0\cr 0&z^{-n}
\end{array}\ri].
\label{OPRHP2}
\eea
Near a hard--edge $z=a$ (where -we recall- $V(z)$ is analytic) we have $\Gamma (z) = [\mathcal O(1), \mathcal O(\ln |z-a|)]$: more precisely we have 
\be
\Gamma(z) =G_a(z) \le[ \begin{array}{cc}
1 & \frac {\kappa(a){\rm e}^{-V(z)}}{2i\pi}\ln(z-a) \\
0&1
\end{array}
\ri]\ ,\ \ \ G_a(z) = \mathcal O(1)\label{OPhardedge}
\ee
Replacing the orthogonality condition (\ref{orthogonality}) by the above jump (and boundary) conditions (\ref{OPRHP2}, \ref{OPhardedge}) we obtain the Riemann-Hilbert problem for the OPs.
\subsection{A problem with constant jumps: ODE}
If we introduce 
\be
\Psi:= \Gamma {\rm e}^{-\frac 1 2 V(z) \sigma_3}
\ee
then we the RHP (\ref{orthogonality}) is turned into  an equivalent RHP
\be
\Psi_+  = \Psi_- \le[
\begin{array}{cc}
1 & \varkappa_j \cr
0&1
\end{array}
\ri]\ ,\ \ \ \ \Psi\sim  (\1 + \mathcal O(z^{-1})) z^{n\sigma_3} {\rm e}^{-\frac {V}2 \sigma_3}
\ee
It is understood that near all singularities of $V(z)$ the matrix $\Psi(z)$ has the singularity implied by the multiplication by $\exp(-\frac V 2\sigma_3)$ and near the hard--edges it has the singularities implied by (\ref{OPhardedge}).

\bp
\label{OPODE}
The matrix $\Psi_n(x)$ solves a rational ODE $\Psi_n'(x)  = D_n(x) \Psi_n(x)$ where $D_n(x)$ is a matrix with rational coefficients with the same singularities as $V'(x)$ and with simple poles at the hard--edges with nilpotent residue.
\ep
{\bf Proof.}
Noticing that $\det \Psi = \det Y \equiv 1$ we have that $\Psi'(z) \Psi^{-1}(z)$ has no jumps across the contours $\Gamma_j$ and hence we have an ODE
\be
\Psi'(z) = D_n(z) \Psi(z)
\ee
where $D_n(z)$ is analytic in the plane punctured at the singularities of $V'$ and at the hard-edges.
\br
The matrix $D_n(z)$ can be written explicitly in terms of the recurrence relations as in \cite{Bertola:Semiiso}. 
\er
Near a singularity $z=a$ of $V'$ we have $\Psi_n(z) = G_a(z) {\rm e}^{-\frac V 2\sigma_3}$ with $G_a(z)$ a formally invertible  asymptotic series in $z-a$ (or $1/z$ if $c=\infty$, in which case $\Psi = G_a (z) z^{n\sigma_3}{\rm e}^{-\frac V 2\sigma_3}$ )  and thus 
\be
D_n(z) =-\frac {V'(z)}2  G(z)^{-1}\sigma_3 G(z)  + \mathcal O(1)
\ee
Near a hard--edge we have 
\be
\Psi_n (z) = 
\wt G_a(z) \le[ \begin{array}{cc}
1 & \frac {\kappa(a)}{2i\pi}\ln(z-a) \\
0&1
\end{array}
\ri]\ ,\ \ \wt G_a = \mathcal O(1)
\ee 
and thus 
\be
D_n(z) = \wt G_a^{-1}(a) \le[
\begin{array}{cc}
0&\frac{\kappa(a)}{2i\pi (z-a)}\\0&0
\end{array}
\ri] \wt G_a(a)  + \mathcal O(1)
\ee

Using Liouville's theorem it is seen that $D(z)$ (which depends on $n$) has the same singularity structure of $V'(z)$ and additionally some simple poles with nilpotent residue at the so--called hard--edges (if any). {\bf Q.E.D.}

\bd
We define  
\be
\Delta_n:= \det [\mu_{i+j}]_{0\leq i,j\leq n-1},\ \ \ \mu_j:= \int_\varkappa x^j {\rm e}^{-V(x)}\d x
\ee
\ed
It is known that $\Delta_n$ is -up to a $n$--dependent constant- the partition function of the Hermitean matrix model \cite{Mehtamethod}.

\bt
The Jimbo-Miwa-Ueno isomonodromic tau function $\tau_n$ associated to the isomonodromic family $\Psi_n$ can be normalized so that 
\be
\tau_n = \Delta_n \tau_0
\ee
with $\tau_0$ given by (\ref{lepalle}) for $V_1 = -V_2 = -\frac 12 V$.
\et
{\bf Proof.}
We show that $\tau_{n+1}/\tau_n  = \Delta_{n+1}/\Delta_n$.

We first observe that $\Psi_{n+1}$ and $\Psi_n$ are related by an elementary Schlesinger transformation as in Sect. \ref{Schlesinger}, namely
\be
\Psi_{n+1}= \le[\begin{array}{cc}
x-c_n &- a_n\\
\frac 1{a_n} & 0
\end{array}
\ri] \Psi_n\label{ladder}
\ee
where $c_n$ and $a_n$ are uniquely determined in terms of the matrices $Y_{\ell}$ that appear in (\ref{OPRHP2}). The relation (\ref{ladder}) is nothing but a rephrasing of the three-term recurrence relation 
\be
x p_n(x)  = p_{n+1} (x) + c_n p_n(x) + a_n p_{n-1}(x)\ .
\ee
From the representation (\ref{OPRHP1}) we have that 
\bea
Y_{1,_{12}}  = \hbox{ coefficient of $z^{-n-1}$ in } \frac 1{2i\pi} \int_\varkappa \frac {p_n(x) {\rm e}^{-V(x)}\d x}{x-z} = -
\frac 1{2i\pi} z^{-n-1} \int_{\varkappa} p_n(x) x^{n} {\rm e}^{-V(x)}\d x
\eea
Since the monic polynomials admit the determinantal representation 
\be
p_n(x) = \frac 1{\Delta_{n}} \det \le[
\begin{array}{cccc}
\mu_0 & \mu_1 & \dots & \mu_{n}\\
\mu_1 & \dots && \mu_{n+1}\\
\vdots &&&\vdots\\
\mu_{n-1}& \dots & &\mu_{2n-1}\\
1 & x& \dots & x^n
\end{array}
\ri]
\ee
we see immediately that $\int_\varkappa p_n(x) x^n {\rm e}^{-V(x)} \d x=  \frac {\Delta_{n+1}}{\Delta_n}
$.

By Thm. \ref{tauratio} and formula (\ref{taudets}) we can choose the normalizations of $\tau_{n+1}$ and $\tau_n$ so that 
\be
\frac {\tau_{n+1}} {\tau_n} = -2i\pi (Y_1)_{12}  = \frac{\Delta_{n+1}}{\Delta_n}\ .
\ee
This immediately implies that $\tau_n = \tau_0 \Delta_n$ (since -by definition- $\Delta_0=1$).

The isomonodromic function $\tau_0$ is associated to the  isomonodromic family $\Psi_0$,
 It is straightforward to verify that the solution of the RHP (\ref {OPRHP2}) for $n=0$ is given by 
\be
\Gamma_0(z) = \le[
\begin{array}{cc}
1 & \frac 1{2i\pi} \int_\varkappa \frac { {\rm e}^{-V(x)} \d x}{x-z}\\
0& 1 
\end{array}
\ri] \ \ \Rightarrow \ \ \Psi_0 (z) :=  \le[
\begin{array}{cc}
{\rm e}^{-\frac {V(z)}2}  & \ds \frac {{\rm e}^{\frac {V(z)}2}} {2i\pi} \int_\varkappa \frac { {\rm e}^{-V(x)} \d x}{x-z}\\
0& {\rm e}^{\frac {V(z)}2}
\end{array}
\ri] 
\ee
A straightforward computation yields for $D_0(x)$
\be
\Psi_0'(z) \Psi_0^{-1}(z) = \le[
\begin{array}{cc}
-\ds\frac {V'(z)}2 &\star\\
0 &\ds \frac {V'(z)}2
\end{array}
\ri] =: D_0(x)
\ee
where the star denotes an expression which is irrelevant to the present considerations (but can be easily computed).  We are thus in the setting of Sec. \ref{sec_trivial}, and thus $\tau_0$ is given by formula (\ref{lepalle}). {\bf Q.E.D.}
\section{Shifted Toeplitz determinants: biorthogonal Laurent and Szeg\"o\ polynomials.}
\label{Toeplitz}
We assume that the moments $\mu_j$ are defined for all $j\in \Z$ and introduce the notation 
\bea
\Delta_n^\ell :=\det\pmatrix {\mu_{\ell} & \mu_{\ell+1} & \cdots &\mu_{\ell+n-1}\cr
\mu_{\ell-1} & \mu_\ell & \cdots & \mu_{\ell+n-2}\cr
 & \ddots & \ddots & \cr
\mu_{\ell-n+1} & \mu_{\ell-n+2} & \cdots & \mu_{\ell}}
\ \ 
\wp_n^\ell(x):= \det\pmatrix {\mu_{\ell} & \mu_{\ell+1} & \cdots &\mu_{\ell+n}\cr
\mu_{\ell-1} & \mu_\ell & \cdots & \mu_{\ell+n-1}\cr
 & \ddots & \ddots & \cr
\mu_{\ell-n+1} & \mu_{\ell-n+2} &\cdots & \mu_{\ell+1}\cr
1& x & \cdots & x^n}\\
\Delta_0^\ell\equiv 1 \ ,\ \ \Delta_{-n}^{\ell} \equiv 0\nonumber 
\eea

It is apparent from the determinantal expression  that $\wp _n^\ell$ is ``orthogonal'' to all powers $x^{\ell-n+1},\dots, x^\ell$
Define the {\bf monic} polynomials as 
\be
p_n^\ell := \frac {\wp_n^\ell}{\Delta_n^\ell} = x^n + \mathcal O(x^{n-1})
\ee

Consider the matrix 
\bea
\Gamma_n^\ell(x) := \le[
\begin{array}{cc}
p_n^\ell  &  \mathcal C[p _n^\ell]\\
c_n^\ell \wp_{n-1}^{\ell-1}  & c_n^\ell x^{-\ell+n- 1} \mathcal C[\zeta ^{\ell-n+1} \wp_{n-1}^{\ell-1} ]
\end{array}
\ri] \ ,\qquad
\mathcal C[f]:= \frac 1 {2i\pi} \int_\kappa \frac{f(z) \mu(z)\d z}{z-x}\\
c_n^\ell := \frac{2i\pi(-1)^{n-1}}{\Delta_n^\ell}
\eea
We have the RHP
\be
\Gamma_+ = \Gamma_- \le[\
\begin{array}{cc}
1 &  \kappa \mu(z)\\
0&1
\end{array}
\ri]\ , \ \ \ \Gamma \sim \le\{
\begin{array}{c}
(\1 + \mathcal O(x^{-1}) \pmatrix{x^{n} & 0 \cr
0& x^{-\ell-1} }\\[18pt]
G_n^\ell (\1 + \mathcal O(x)) \pmatrix { 1 & 0 \cr
0&x^{n-\ell-1} }
\end{array}
\ri.
\ee
where 
\be
G_n^\ell = \le[
\begin{array}{cc}
\frac {(-)^n \Delta_{n}^{\ell+1}}{\Delta_{n}^\ell} &  \frac {- \Delta_{n+1}^\ell}{2i\pi \Delta_n^\ell} \\
-\frac {2i\pi \Delta_{n-1}^\ell}{\Delta_n^\ell} & \frac {(-1)^{n+1} \Delta_{n}^{\ell-1}}{\Delta_{n}^\ell}
 \end{array}
\ri]\ ,\ \ \det G_n^\ell = \frac {\Delta_{n+1}^\ell \Delta_{n-1}^\ell  +  \Delta_n^{\ell-1} \Delta_n^{\ell+1}}{(\Delta_n^\ell)^2} = 1 
\label{LOPs}
\ee
Note that 
\be
\det \Gamma_n^\ell (x) = x^{n-\ell-1}
\ee
\subsection{Constant-jump RHP and ODE}
We now postulate $\mu(x) = {\rm e}^{-V(x)}$ with $V'(x)$ a rational function.
We introduce the new matrix 
\bea
\Psi &\&= \Psi_n^\ell:= \Gamma_n^\ell {\rm e}^{-\frac 12 V(x) \sigma_3}\\
\Psi_+ &\& = \Psi_-\le[
\begin{array}{cc}
1 & \kappa\\
0&1
\end{array}
\ri]\ , \cr 
\Psi&\&  = \le(\1 + \mathcal O(x^{-1})\ri) \pmatrix{ x^n & 0 \cr
0& x^{-\ell-1}} {\rm e}^{-\frac 12 V(x) \sigma_3}\cr
\Psi &\&= \le(\1 + \mathcal O(x)\ri) \pmatrix{ 1 & 0 \cr
0& x^{n-\ell-1}} {\rm e}^{-\frac 12 V(x) \sigma_3}
\eea
This matrix solves a similar RHP with constant jumps and hence also an ODE
\be
\pa_x \Psi_n^\ell  = D_n^\ell(x) \Psi_n^\ell
\ee
As in Prop. \ref{OPODE} one can see that the connection has the same singularities as
$\wh V'(x)$, where  
\be
\wh V(x):= V(x) - (n-\ell -1) \ln x\ .\label{shiftV}
\ee
plus nilpotent-residue simple poles at the hard--edges.

The various shifts $n\to n \pm 1$ and $\ell\mapsto \ell\pm 1$ all correspond to Schlesinger transformations (of more general form than the one used for OP). In \cite{Bertola:LOPs} we considered (from a different approach) the shifts 
\be
\begin{array}{cccc}
(n,\ell) &\mapsto& (n+1,\ell) & \hbox {\bf Circle move}\\
(n,\ell)&\mapsto &(n+1, \ell+1) & \hbox{\bf Line move}\\
(n,\ell)& \mapsto &(n,\ell+1) & \hbox {\bf Circle-to-line move}
\end{array}
\ee
Of course the ``Line move'' is simply the composition of the other two; the reason for the naming was that a circle move implies that the OP satisfy a recurrence of the form
\be
x(p_n^\ell + \star p_{n-1}^{\ell-1}) = p_{n+1}^\ell + \star p_n^\ell 
\ee
for some constants indicated anonymously by the  $\star$. This recurrence is typical of the Szeg\"o\ orthogonal polynomials on the circle, whence the name.

Likewise a line move corresponds to a recurrence of the form
\be
xp_n^\ell  = p_{n+1}^{\ell+1} + \star p_n^\ell  + \star p_{n-1}^{\ell-1}
\ee
typical of the recurrence relation for orthogonal polynomials on the real line. The circle-to-line move is yet a different recurrence that intertwines the two. 
\br
These mixed recurrence relations were important in the study of integrable lattices \cite{Gekhtman:InverseMoment, Gekhtman:Elementary} and put on the same grounds the integration of all known lattices (Toda, relativistic Toda, Ablowitz-Ladik, Volterra) and also some unnamed generalizations. 
\er
\br
If all the moves are Circle-moves, then the resulting OPs are Szego polynomials on the unit circle (if the  weight is a real weight on the circle, that is, or generalizations thereof). If all the moves are Line-moves then the polynomials are the usual (non Hermitean) orthogonal polynomials. 
\er
Following Thm. \ref{tauratio} and formula \ref{taudets} together with the expressions (\ref{LOPs})  one can verify that -in all cases-
\be
\frac {\Delta_{n'}^{\ell'}}{\Delta_n^\ell}   = \frac {\tau_{n'}^{\ell'}}{\tau_n^\ell}\label{taufrac}
\ee
Indeed a Schlesinger transformation that changes the exponents only at infinity (like in Sect. \ref{Hankel}) corresponds to a Line-Move and $\tau' /\tau$ is given by the $(1,2)$ entry of the $Y_1$ term in the expansion at infinity.
The other moves are Schlesinger transforms that change one exponent at $\infty$ and another exponent of formal monodromy at $x=0$. In particular (referring to (\ref{LOPs}))
\begin{enumerate}
\item for a Circle Move, $\frac {\tau_{n+1}^\ell}{\tau_n^\ell}  \propto  (G_n^\ell)_{12}$;
\item for a Circle-to-Line Move, $\frac {\tau_{n}^\ell}{\tau_{n}^{\ell+1} }\propto (G_n^\ell)_{11}$. 
\end{enumerate}
Once more $\propto$ means up to any expression that does not depend on the isomonodromic parameters. Inspection of $G_n^\ell$ yields (\ref{taufrac}).

%

The simplest way to completely describe the relation between $\Delta_n^\ell$ and $\tau_n^\ell$ is to note that the shifted T\"oplitz determinant is (up to a reshuffling of rows) a H\"ankel determinant of the same size for the weight  $x^{n-\ell-1} \mu(x)\d x$. This corresponds to considering the isomonodromic equation for 
\be
\wh \Psi_n^\ell (x) := \Psi_n^\ell(x)  x^{\frac{\ell-n+1}2}
\ee
\bea
\wh \Psi_n^\ell(x)_+ &\& = \wh \Psi_n^\ell(x)_-\le[
\begin{array}{cc}
1 & \kappa\\
0&1
\end{array}
\ri]\ , \cr 
\wh\Psi_n^\ell &\&  = \le(\1 + \mathcal O(x^{-1})\ri) x^{n\sigma_3} {\rm e}^{-\frac 12 \wh V(x) \sigma_3}\ \ \ 
\wh \Psi_n^\ell = G_n^\ell \le(\1 + \mathcal O(x)\ri) {\rm e}^{-\frac 12 \wh V(x) \sigma_3}
\eea
which shows that --as the reader may have observed--  the polynomials $\wp_n^\ell$ and $\wp_{n-1}^{\ell-1}$ are simply the (unnormalized) orthogonal polynomials for the weight $x^{n-\ell-1} \mu(x)\d x$ (on the same contours $\kappa$).

This implies that 
\be
\tau_n^\ell = \tau_0 \Delta_n^\ell
\ee
where $\tau_0$ is the same expression in (\ref{lepalle}) but with $V_1 = -V_2 = -\frac 1 2 \wh V$.
This result reproduces the results of \cite{Bertola:LOPs} for shifted T\"opliz determinants and \cite{Its:RandomWords2} for (unshifted) T\"oplitz determinants with special symbol.

\section{Bimoment determinants: Cauchy biorthogonal polynomials}
\label{Cauchy}
Let $\a(x), \beta(y)$ be two arbitrary semiclassical weights and consider the bi-moment functional 
\be
\mathcal I_{ij}:=  \int_{\kappa_y} \int_{\kappa_x} x^i y^j \frac {\a(x) \beta(y)\d x \d y}{x+y}
\ee
where the symbol $\kappa_{x,y}$ denotes any linear combination of contours as long as the integrals make sense. It is understood that the contours of integration for the weights  $\a$ and $\beta$ must be such that the intersections of the contours $\kappa_x\cap (-\kappa_y)$ consist of isolated points, so that the integral is well defined.

We denote by $\Delta_n$ the principal minors of size $n$
\be
\Delta_n:= \det [I]_{0\leq i,j\leq n-1} \label{bimomdet}
\ee

The {\bf biorthogonal polynomials} are two sequences of (monic) polynomials $\{p_n(x)\}_{n\in \N}, \{q_n(y)\}_{n\in \N}$ of exact degree $n$ such that 
\be
\int_{\varkappa_y } \int_{\varkappa_x}  p_n(x) q_m(y) \frac {\a(x) \beta(y)\d x \d y}{x+y} =  \frac{ \Delta_{n+1}}{\Delta_n} \delta_{nm}
\ee  
(Note that the form of the constant appearing in the RHS is actually a small theorem).

They admit the determinantal representation
\be
p_n(x) = \frac 1{\Delta_n} \det \le[
\begin{array}{ccc}
\mathcal I_{00}&\dots& \mathcal I_{0 n}\\
&&\\
\mathcal I_{n-1\, 0}& \dots & \mathcal I_{n-1\,n}\\
1 & \dots & x^n
\end{array}
\ri]\ ,\ \ 
q_n(y) = \frac 1{\Delta_n} \det \le[
\begin{array}{cccc}
\mathcal I_{00}&\dots& \mathcal I_{0 \,n-1} & 1\\
&& & \vdots\\
\mathcal I_{n\, 0}& \dots & \mathcal I_{n\,n-1} & y^n
\end{array}
\ri]
\ee
\br
If $\a,\beta$ are positive measures and the contour of integration are the positive real axis then it was shown in \cite{Bertola:CauchyMM} that the bimoment matrix is  {\bf totally positive}; moreover the zeroes of the OPs are {\bf real, simple and interlaced}.  
It was also shown ibidem the these BOPs satisfy a four-term recurrence relation\footnote{It was proven for positive weights on the real axis, but the proof is entirely algebraic and hence applies without changes to this semiclassical situation.}.
\er

In \cite{Bertola:CBOPs} the BOPs and some auxiliary polynomials were introduced in order to  describe the Christoffel-Darboux identities, which in turn have important applications to the study of the spectral statistics for the Cauchy matrix model \cite{Bertola:CauchyMM}.  In the present work we will only need the following characterization for the sequence of (monic) OPs $p_n(x)$ (with an entirely specular characterization for the $q_n(y)$).

\bp[Prop. 8.2 in \cite{Bertola:CBOPs}]
\label{RHP2}

Consider the Riemann--Hilbert problem (RHP$_n$) of finding a matrix $\wh \Gamma(z) = \wh \Gamma_n(z)$ such that 
\begin{enumerate}
\item $\wh \Gamma(z)$ is analytic on $\C \setminus (\kappa_x\cup (-\kappa_y))$
\item $\wh \Gamma(z)$ satisfies the jump conditions 
\bea
 \label{eq:RHphat}
\wh \Gamma(z) _+ = \wh \Gamma(z)_-
 \le[
\begin{array}{ccc}
1 &   \kappa_x \alpha(z)  & 0\cr
0&1&0\\
0&0&1
\end{array}
\ri]\ ,\ \ z\in \kappa_x\\
\
\wh \Gamma(z) _+ = \wh \Gamma(z)_- \le[
\begin{array}{ccc}
1 &  0 & 0\cr
0&1& \kappa_y  \beta^\star(z)\\
0&0&1
\end{array}
\ri]\ ,\ \ z\in -\kappa_y , \\
\beta^\star(z):= \beta(-z)\nonumber
\eea
\item its asymptotic behavior at $z=\infty$ $\Im(z)\neq 0$ is 
\bea \label{eq:Gammahat-as}
\wh \Gamma(z) =  \le(\1  + \mathcal O\le(\frac 1 z\ri)\ri)\le[
\begin{array}{ccc}
z^n & 0&0 \cr
0 &1 &0\cr
0 &0& \frac 1{ z^{n}} 
\end{array} 
\ri].  \label{RHhatasymp}
\eea

\end{enumerate}
Then such a $\wh \Gamma(z)$  uniquely characterizes the polynomial $p_n(z)$ as its $(11)$-entry. Moreover $\wh \Gamma(z)$ can equivalently be written as:
\bea
\label{eq:p-recovery}
\wh \Gamma(z)=
\le[\begin{array}{ccc}
p_{0,n}&p_{1,n}&p_{2,n}\\
\wh p_{0,n-1}&\wh p_{1,n-1}&\wh p_{2,n-1}\\ \wt p_{0,n-1}& \wt p _{1,n-1}&\wt p_{2,n-1} 
\end{array}\ri]
 \eea
where the first column consists of polynomials of the indicated degree and, for any function $f(z)$ we have denoted
\bea
f_0(z):= f(z)\ ,\qquad f_1(z):=\frac 1{2i\pi} \int_{\kappa_x} \frac {f(x) \a(x)\d x }{x-z}\ ,\qquad
f_2(z):= \frac 1{2i\pi}\int_{\kappa_y} \frac {f_1(-y) \beta(y) \d y}{z+y}\label{Ctran}
\eea
\ep
\br
The detailed description for the entries of $\Gamma$ (although for a different normalization) can be found in \cite{Bertola:CBOPs}.
\er
Since in \cite{Bertola:CBOPs} the proposition was stated only for positive measures on the real axis, we briefly sketch the proof that the above RHP characterizes the OP's $p_n$. 
The jump-relations and the asymptotics at infinity imply that the first column is an entire function bounded by $z^n$ and hence -by Liouville's theorem- a polynomial. Moreover the asymptotic at $\infty$ imply that the $(11)$ entry is a monic polynomial of degree $n$ and the remaining entries some polynomials of lesser degree.

The jump relations imply that the second and third column are obtained from the first by the expressions (\ref{Ctran}) (using Sokhotsky-Plemelj formula). Finally, the $(33)$ entry of the asymptotic at infinity implies that 
\be
\int_{\kappa_y}\!\!\int_{\kappa_x} p_n(x)y^j \frac {\a(x)\beta(y) \d x \d y}{x+y} = 0\ ,\ \ j\leq n-1
\ee
which is the orthogonality requirement.

\subsection{Constant-jump RHP and ODE}
A simple transformation brings the above RHP into one with constant jumps. 
We will write $\a(x) = {\rm e}^{-V_1(x)}$ and $\beta^\star(x):= \beta(-x) = {\rm e}^{-V_2(x)}$. We introduce the new matrix 
\be
\Psi(z):= \wh \Gamma(z)  \le[
\begin{array}{ccc}
\exp\le(-\frac {2V_1 + V_2}3 \ri) & &\\
& \exp\le(\frac {V_1-V_2}3\ri)& \\
&& \exp\le(\frac {2V_2+ V_1}3\ri)
\end{array}
\ri]
\ee
which solves a RHP with constant jumps 
\be
\Psi(z) _+ = \Psi(z)_-
 \le[
\begin{array}{ccc}
1 &    \kappa_x & 0\cr
0&1&0\\
0&0&1
\end{array}
\ri]\ ,\ \ z\in \kappa_x\\
\
\Psi(z) _+ = \Psi(z)_- \le[
\begin{array}{ccc}
1 &  0 & 0\cr
0&1& \kappa_y\\
0&0&1
\end{array}
\ri]\ ,\ \ z\in -\kappa_y , 
\ee
In the same way as for the OP we can prove
\bp
The matrix $\Psi = \Psi_n(z)$ solves an  ODE of the form 
\be
\Psi_n'(z)  = D_n(z) \Psi_n(z)
\ee
where the $3\times 3$ matrix $D_n(z)$ has the same singularities as $V_1'(z), V_2'(z)$ plus simple poles (with nilpotent residues) at the hard-edges.
\ep

\bp
\label{Cauchyratio}
The ratio of two consecutive tau functions satisfy 
\be
\frac{\tau_{n+1}}{\tau_n}= \frac {\Delta_{n+1}}{\Delta_n}
\ee
\ep
{\bf Proof.}
By inspection of (\ref{RHhatasymp}) we observe  that the shift $\Psi_{n} \mapsto \Psi_{n+1}$ is  a Schlesinger transformation as in Sect. \ref{Schlesinger} involving the exponents of formal monodromy at infinity in the entries $(1,1)$ and $(3,3)$.
Once again, using Thm. \ref{tauratio} and formula (\ref{taudets}) we have (with a convenient normalization of tau-functions)
\be
\frac {\tau_{n+1}}{\tau_n} =(-)^n Y_{1,_{1\,3}}
\ee
It is immediate to see that $(Y_{1})_{1\,3}$ is the coefficient in front of $z^{-n-1}$ in the asymptotic expansion of $p_{2,n}$ namely
\bea
p_{2,n}(z) =  - \int_{\kappa_y}\int_{\kappa_x}  p_n(x) \frac {\a(x) \beta(y)\d x \d y}{(z+y)(y+x)}  = (-)^n z^{-n-1} \int_{\kappa_y}\int_{\kappa_x}  p_n(x) y^n \frac {\a(x) \beta(y)\d x \d y}{y+x}   + \mathcal O(z^{-n-2}) =\\ 
= (-)^n z^{-n-1} \frac {\Delta_{n+1}} {\Delta_n}  + \mathcal O(z^{-n-2})
\eea
This concludes the proof. {\bf Q.E.D.}\par \vskip 5pt

We thus prove 
\bt
We can choose the normalization for the Jimbo-Miwa-Ueno tau functions $\tau_n$ for the isomonodromic family $\Psi_n$ so that it is related to the bimoment determinant $\Delta_n$  (\ref{bimomdet}) by 
\be
\tau_n = \tau_0 \Delta_n
\ee
with $\tau_0$ given by (\ref{lepalle}) with $V_1 = -\frac {2V_1 + V_2}3,\  V_2 = \frac {V_1-V_2}3 \ ,\ V_3 =  \frac {2V_2+ V_1}3$ (here the $V_{1,2,3}$ on the LHS refer to the $V_j$'s appearing in (\ref{lepalle}), while $V_{1,2}$ on the RHS refer to the potentials used in the definition of the weights).
\et
{\bf Proof.} By Prop. \ref{Cauchyratio} we already know $\tau_n = \tau_0 \Delta_n$, so it only remains to compute $\tau_0$. The RHP$_0$ for $\Psi_0$ is easily solvable and yields
\be
\Psi_0  = \le[
\begin{array}{ccc}
1 & \frac 1{2i\pi} \int_{\kappa_x} \frac {\a \d x}{x-z} &  \frac 1{4\pi^2} \int_{\kappa_y}\int_ {\kappa_x} \frac {\a(x)\beta(y)\d x \d y}{(z+y)(y+x)}\\
&1 & \frac 1{2i\pi} \int_{\kappa_y} \frac {\beta \d y}{y-z}\\
&&1
\end{array}
\ri]\le[
\begin{array}{ccc}
\exp\le(-\frac {2V_1 + V_2}3 \ri) & &\\
& \exp\le(\frac {V_1-V_2}3\ri)& \\
&& \exp\le(\frac {2V_2+ V_1}3\ri)
\end{array}
\ri]
\ee
Thus the ODE $\pa_z - D_0(z)$ has an upper--triangular form 
\be
D_0(z) = \le[
\begin{array}{ccc}
 - \frac {2V_1'+V_2'}3 & \star & \star\\
 & \frac {V_1'-V_2'}3 & \star\\
 && \frac {2V_2' + V_1'}3
\end{array}
\ri]
\ee
where the precise expressions for the upper--triangular entries is presently irrelevant.
We are thus in the setting of Sec. \ref{sec_trivial} and we can use eq. \ref{lepalle}. {\bf Q.E.D.}

\br 
We remark -in passing - that the spectral curve of $D_0$ (i.e. its characteristic polynomial) is a completely factorized rational curve
\be
 \le(\lambda +  \frac  {2V_1'+V_2'}3\ri) \le(\lambda - \frac  {V_1'-V_2'}3\ri) \le( \lambda - \frac {2V_2' + V_1'}3 \ri) =0 
\ee
\er
\section{Multi-measure moment matrices: multiple orthogonal polynomials}
\label{Multimeas}
Multiple orthogonal polynomials appear in the theory of simultaneous Pad\'e\ approximations (for a review and interesting applications to number-theory and spectral theory of banded matrices see \cite{Aptekarev:Multiple}).

We will give a (unfairly) quick definition which allows us to get to the point as directly as possible.
Consider $K$ measures $\mu^{(j)}(x)$. 
Denote by $\vec n \in \N^K$ a multi-component vector and $|\vec n| = \sum_{j=1}^K n_j$ \cite{VanAssche:RHPMOPs}.

The multiple orthogonal polynomials of type II are polynomials of degree $|\vec n|$ defined by 
\be
\int_{\kappa_j} P_{\vec n}(x) x^\ell \mu^{(j)} (x) \d x = 0 \ ,\ \  \forall \ell:\ 0\leq \ell \leq n_j-1, j=1,\dots K.
\ee
The system is called {\bf perfect} if $P_{\vec n}$ has exact degree $n := |\vec n|$.
We can then assume that $P_{\vec n}$ is {\bf monic} and the following determinantal expression are  immediately deduced
\be
P_{\vec n}(x) = \frac 1{\Delta_{\vec n}} \det \le[
\begin{array}{ccccccc}
\mu^{(1)}_0 &&&&&& \mu_{n}^{(1)}\\
\mu^{(1)}_1 &&&&&& \mu_{n+1}^{(1)}\\
\vdots &&&&&& \vdots\\
\mu^{(1)}_{n_1-1} &&&&&&\mu_{n+n_1-1}^{(1)}\\
\hline
\vdots &&&&&\vdots\\
\hline
\mu^{(K)}_0 &&&&&& \mu_{n}^{(K)}\\
\mu^{(K)}_1 &&&&&& \mu_{n+1}^{(K)}\\
\vdots &&&&&& \vdots\\
\mu^{(K)}_{n_K-1} &&&&&&\mu_{n+n_K-1}^{(K)}\\
\hline 
1& x&\dots &&&&x^n
\end{array}
\ri]\\
\Delta_{\vec n}:= \det \le[
\begin{array}{ccccccc}
\mu^{(1)}_0 &&&&&& \mu_{n-1}^{(1)}\\
\vdots &&&&&& \vdots\\
\mu^{(1)}_{n_1-1} &&&&&&\mu_{n+n_1-2}^{(1)}\\
\hline
\vdots &&&&&\vdots\\
\hline
\mu^{(K)}_0 &&&&&& \mu_{n-1}^{(K)}\\
\vdots &&&&&& \vdots\\
\mu^{(K)}_{n_K-1} &&&&&&\mu_{n+n_K-2}^{(K)}\\
\end{array}
\ri]
\label{multidet}
\ee
They are characterized as the $(1,1)$ entry of the solution of the following Riemann--Hilbert problem
\bea
\Gamma_{\vec n}(x)_+ = \Gamma_n(x)_- \le[
\begin{array}{cccc}
1 & \kappa_1 \mu^{(1)} & \dots  &\kappa_K \mu^{(K)}\\
 & 1&&\\
&&\ddots&\\
&&&1
\end{array}
\ri]=: \Gamma_n(x)_{_-} M(x)\cr
\Gamma_n(x) = (\1 + \mathcal O(x^{-1})) \le[
\begin{array}{cccc}
x^n &&&\\
& x^{-n_1} &&\\
&&\ddots &\\
&&& x^{-n_K}
\end{array}
\ri]\label{multiRHP}
\eea
We observe that the first row of the solution $\Gamma  = \Gamma_{\vec n}$ is written explicitly as 
\be
\Gamma_{11} = P_{\vec n}(x)\ ,\qquad 
\Gamma_{1j} =\frac 1{2i\pi} \int_{\kappa_j} \frac{P_{\vec n}(\xi) \mu^{(j)}(\xi)}{\xi-x} \d \xi
\ee
\subsection{Constant jump RHP and ODE}
As  it stands the RHP (\ref{multiRHP}) cannot be related directly to the theory of isomonodromic deformations, since the latter refers to deformations of rational ODEs. For general weights $\mu^{(j)}$ the RHP cannot be reduced to a rational ODE. 

In order to reduce the problem to one with constant jumps it is sufficient to find a matrix $G(x)$ such that $G(x)_-^{-1} M(x) G(x)_+$ is (locally) constant. 

The properties of $G(x)$ depend on the class of weights we are considering. We will only consider semiclassical weights but there are situations (which we will consider elsewhere) in which the weights are not semiclassical and yet a more refined approach can be used to obtain a rational ODE.
\subsubsection{Semiclassical weights}
If the weights are semiclassical and of the form $\mu_\ell = {\rm e}^{-V_\ell}$ we can chose $G(x)$ as follows
\bea
G(x) = {\rm diag} \bigg[{\rm e}^{ - \wh V_0} , {\rm e}^{\wh V_1(x)}, \dots, {\rm e}^{\wh V_K(x)}\bigg]\\
\wh V_0:= \sum_{j=1}^K V_j\ ,\qquad 
\wh V_j:= \wh V_0 -V_j  = \sum_{k\neq j} V_k\ .
\eea
We thus define 
\be
\Psi = \Psi_{\vec n} = \Gamma_{\vec n}(x) G(x)
\ee
which now solves a constant-jump RHP and following the same steps as Prop. \ref{OPODE} we obtain an ODE of the form 
\be
\Psi_{\vec n}'(x) = D_{\vec n}(x) \Psi_{\vec n}(x)
\ee
where $D_{\vec n}(x)$ is a rational matrix with the same singularity structure as the (totality of the) $V_j'$s as well as nilpotent-residue simple poles at the hard-edges (if any).

\bt
The multi-measure determinant $\Delta_{\vec n}$ in eq. (\ref{multidet}) and the Jimbo-Miwa-Ueno tau function $\tau_{\vec n}$  for the isomonodromic family $\Psi_{\vec n}$ are related by 
\be
\tau_{\vec n}  = \Delta_{\vec n} \tau_{\vec 0}\ ,
\ee
with $\tau_{\vec 0}$ given by formula (\ref{lepalle}) with $V_1 = -\wh V_0$ and $V_\ell = \wh V_\ell\ ,\ \ell = 2,\dots K+1$.
\et
{\bf Proof.}
The matrix $\Psi_{\vec n}$ has the asymptotics
\bea
\Psi_{\vec n}(x) = \le(\1 +\frac {Y_1}x + \frac {Y_2}{x^2} + \dots \ri) x^{T_{\vec n}} G(x)\cr
T_{\vec n } = {\rm diag} (n,-n_1,\dots, -n_K)\ .
\eea

Suppose $\vec n'  = \vec n + e_\ell$, where $e_\ell$ is an standard basis vector. The transformation $\vec n\mapsto \vec n'$ correspond to a Schlesinger transformation with (in the notation of eq. (\ref{sigmaij})) $\sigma_{1,1+\ell}$. 
Then Thm. \ref{tauratio} and formula (\ref{taudets}) state that  
\be
\frac {\tau_{\vec n'}}{\tau_{\vec n}} = Y_{1,_{1,\ell+1}}
\ee 
and inspection of the matrix $\Gamma_{\vec n}$ yields 
\be
Y_{1,_{1,\ell+1}} \propto  \int_{\kappa_\ell} P_{\vec n}(x) x^{n_\ell} \mu^{(\ell)} (x)\d x \propto \frac {\Delta_{\vec n'}}{\Delta_{\vec n}}
\ee
where the proportionality is within signs or other constant factors that are irrelevant because of the inherent ambiguity in the definition of isomonodromic tau function.

With this we have proved that the ratio any two adjacent multi-measure moment determinants is the same as the ratio of the corresponding isomonodromic tau functions
\be
 \frac {\Delta_{\vec n'}}{\Delta_{\vec n}} = \frac {\tau_{\vec n'}}{\tau_{\vec n}} 
\ee
and thus 
\be
\Delta_{\vec n} \tau_{\vec 0} =\tau_{\vec n} 
\ee
Once more, the computation of $\tau_{\vec 0}$ is simple because $D_{\vec 0}$ is an upper-triangular connection 
\bea
\Psi_{\vec 0}(x) = \le[
\begin{array}{ccc}
1 & W_1(x) & \dots W_K(x)\\
&\ddots &\\
&&1
\end{array}
\ri]G(x)\ ,\qquad W_j(x):= \frac 1{2i\pi} \int_{\kappa_j} \frac {\mu^{(j)}(\xi) \d \xi}{\xi-x}\\
D_{\vec 0} = \le[
\begin{array}{cccc}
-\wh V_0'(x) &\star &\star&\star \\
&\wh V_1'(x) &\\
&&\ddots&\\
&&&\wh  V_K'(x)
\end{array}
\ri]
\eea
Once more we are in the setting of Sec. \ref{sec_trivial} and formula (\ref{lepalle}) applies. {\bf Q.E.D.}
\section{Conclusion}
It is apparent that the proofs for the  four classes of moment determinants are essentially a repetition of the same idea. In order not to overburden the paper we did not consider other cases that nevertheless might have interesting applications. These include (but are not necessarily limited to) the cases of Angelescu and Nikishin systems for semiclassical weights. In fact the case of Sec. \ref{Cauchy} is an instance of Nikishin system \cite{Aptekarev:Multiple} but we singled it out because of its independent appearance  in the context of random matrix theory \cite{Bertola:CauchyMM}. 

Another important multimatrix model that relates to biorthogonal polynomials is the Itzykson--Zuber matrix model with a partition function 
\be
Z_{n}^{IZ}:= \int \d M_1 \d M_2 {\rm e}^{-{\rm Tr} V_1(M_1) - {\rm Tr} V_2(M_2) + {\rm Tr} M_1M_2}
\ee
which corresponds to the following bimoment determinant \cite{EynardMehta, Bertola:DualityCMP}
\be
Z_{n}^{IZ}\propto \det{\mathcal I_{jk}}\ ,\ \ \ \mathcal I_{jk} = \int \int \d x \d y x^j y^k{\rm e}^{-V_1(x)-V_2(y) + xy}\ .
\ee
This determinant can also be related to multiple orthogonal polynomials \cite{Kuijlaars:RHPBOPs} but for non semi-classical weights. While the general idea of the proof $Z_n^{IZ}$ is  an isomonodromic tau function is the same, the details are substantially different. The main reason is that the relevant isomonodromic family does not have an asymptotic expansion as in (\ref{exppsi}) at $\infty$ and thus the definition of isomonodromic tau function of the Japanese school cannot be applied as is. In particular the leading coefficient of the relevant ODE has a very degenerate spectrum and the asymptotics involves fractional powers of $z$ \cite{Bertola:DifferentialCMP, Bertola:Isomono_resonant}. The proof in this case will be detailed in a forthcoming paper \cite{Bertola:IZtau}.

\bibliographystyle{plain}
\bibliography{/Users/bertola/Documents/Papers/BibDeskLibrary}
\end{document}